\documentclass[ journal]{IEEEtran}  
\usepackage{amsfonts}
\usepackage{xcolor}
\usepackage{amsmath,graphicx,url,times}
\usepackage{cite}
\graphicspath{{figures/}}

\begin{document}
\title{Directional emphasis in ambisonics} 
\author{W. Bastiaan Kleijn
\thanks{W. B. Kleijn is with Google LLC, San Francisco and Victoria University of Wellington,
email: see http://ecs.victoria.ac.nz/Main/BastiaanKleijn}}
\maketitle 
\begin{abstract}
 We describe an ambisonics enhancement method that increases the signal strength in specified
 directions at low computational cost. The method can be used in a static setup to emphasize
 the signal arriving from a particular direction or set of directions. It can also be used in an adaptive arrangement
 where it sharpens directionality and reduces the distortion in timbre associated
 with low-degree ambisonics representations. The emphasis operator has very low
 computational complexity and can be applied to time-domain as well as time-frequency
 ambisonics representations. The operator upscales a low-degree  ambisonics representation
 to a higher degree representation. 
\end{abstract}

\IEEEpeerreviewmaketitle
\section{Introduction}
\label{s:intro}
Ambisonics \cite{gerzon:73,gerzon:85,daniel1998,daniel2003, frank2015producing} is a
representation for sound fields 
that can take the form of a series of countably infinite
spatial basis functions, each multiplied by a temporal scalar audio signal.
Signal acquisition and rate (storage) constraints lead to truncation of the series,
limiting the \textit{degree} of the ambisonics description. A low ambisonics degree
bounds the frequency-dependent radius within which the sound field is described
accurately and may make the soundfield unnatural outside this region.  This can lead to
audible artefacts in a signal rendered from the low-degree 
description, e.g., \cite{daniel1998,daniel2003, zotter2012all,frank2015producing,berge2010high,Wabnitz11, Wabnitz12, heller2012toolkit, Kleijn2017g}. We
describe an approach that addresses these issues and additionally facilitates directional
emphasis for other purposes.

The artefacts of low-degree ambisonics can be explained as follows
\cite{Kleijn2017g}. Conventional rendering 
uses the Moore-Penrose inverse to map the ambisonics representation to
loudspeaker signals. Hence it minimizes the energy produced by the virtual or
physical loudspeakers subject 
to the known ambisonics coefficients being correct. The constraint means the sound
field is accurate in a \textit{sweet} zone around the origin. The imposed energy efficiency
requires the loudspeaker contributions to the sound-field 
to add coherently in the sweet zone, but not outside.  As the radius of the zone is frequency
dependent, a low-pass timbre is heard by a listener at the origin. 
More-over, energy minimization implies the distribution of the acoustic energy over many loudspeakers,
reducing the directionality of the sound field outside the sweet zone.

We define our objective more carefully. Consider a sound field in a source-free region
around the origin of our coordinate system. The sound field in this
region can be generated by a continuous density of monopole sound sources located on a
2-sphere centered at the origin, e.g., \cite{poletti2005three}.  The temporal source
signals of the monopole sources form a two-dimensional (2D) scalar \textit{source field} on the
2-sphere that forms an alternative specification of the sound field.
The strengthening of the directionality of the sound field can then be
defined as the emphasizing of the source field on the 2-sphere. The
emphasis operator can be interpreted as an \textit{acoustic spotlight}.

We aim to implement the emphasis operator directly in the ambisonics representation.  Our
objective is a low-complexity operator that applies to both time domain
and time-frequency domain representations. In addition to the standard goal of
\textit{adaptive emphasis} of the sound field (strengthening existing directionality), a secondary goal is \textit{static emphasis}
of the sound field (time-invariant emphasis operator).

We are not aware of existing systems that provide static emphasis (time-invariant
emphasis) of an ambisonics representation. Adaptive emphasis operators can be
classified into two classes. The first class does not change the ambisonics degree. The
common max $r_E$ emphasis operator \cite{daniel1998,daniel2003} minimizes sidelobes 
resulting from the  truncation to a low degree representation \cite{zotter2012all}.
The second class upscales the low-degree ambisonics information into a high-degree ambisonics 
representation \cite{Wabnitz11, Wabnitz12, Kleijn2017g}.  Only the second  class
facilitates \textit{idempotency}: the analysis of the rendered  sound field returns the original
ambisonics description. 
In addition to the fore-mentioned classes, methods exist that are an integral component
of rendering, e.g., \cite{pulkki2007spatial,berge2010high}, usually
restricted to mapping the ambisonics representation into one or two plane waves
\cite{frank2015producing}. 

Our contribution is an emphasis operator that has two advantages compared to the
state-of-the-art operators for idempotent rendering \cite{Wabnitz11, Wabnitz12,
  Kleijn2017g}.  First, it can be used in both static and adaptive emphasis applications
(existing methods are aimed at adaptive 
emphasis). Second, our operator, which is based on Clebsch-Gordan
coefficients, has low computational complexity. It can be used in the time domain and
it raises the ambisonics degree with a matrix multiplication requiring only a handful
of multiplies per output sample. To ensure idempotent adaptive rendering, the 
operator can incorporate a projection \cite{Kleijn2017g} without added computational cost.


\section{Theory}
\label{s:theory}

This section first describes the source field in section 
\ref{s:relating}, then defines the emphasis operator in section \ref{s:emphasis} and
methods to compute it in section \ref{s:finding}. The discussion is for complex spherical
harmonics but extends to the real case. Similarly to, e.g.,\cite{spors2012}, we write the
spherical harmonics as  
\begin{align}
Y_n^m(\theta, \phi) &= (-1)^m\sqrt{
\frac{(2n+1)}{4 \pi}
\frac{ (n-|m|)!}{(n+|m|)!}}
 P_n^{|m|}(\cos(\theta))  \mathrm{e}^{im\phi} ,
\label{q:sphh}
\end{align}
where the $P_n^{m}$ are the associated Legendre functions,  $\theta\in[0,\pi]$ is
elevation and $\phi\in[-\pi,\pi]$ is the azimuth. The $Y_n^m(\cdot,\cdot)$ are orthonormal
on the unit 2-sphere and use the  Condon-Shortly phase convention.
The form of \eqref{q:sphh} implies that $Y_n^{m*}(\theta,\phi) = Y_n^{-m}(\theta,\phi)$,
simplifying derivations. 

\subsection{Relating the 3D Sound Field and the 2D Source Field}
\label{s:relating}
Our aim in this subsection is to provide the background for deriving an emphasis operator in
section \ref{s:emphasis}. While it is not obvious how to define an emphasis of the sound 
field directly, it is clear that such an emphasis corresponds to a sharpening of the
source field on the 2-sphere defined in section \ref{s:intro}. 

We follow an approach used earlier in \cite{poletti2005three} and
\cite{wu2009theory}. While the approach is illustrated in the
frequency domain, the same reasoning holds in the time domain. We consider an internal
sound field expansion of the form
\begin{align}
p(r,\theta,\phi,k) = \sum_{n=0}^\infty \sum_{m=-n}^n  B_n^m(k)j_n(kr)  Y_n^{m}(\theta,\phi) ,
\label{q:basic}
\end{align}
where $p(\cdot)$ is pressure, $r$ is radius, $j_n(\cdot)$ is the
spherical Bessel function,  $B_n^m$ are the ambisonics coefficients and
$k=\frac{\omega}{c}$ is the wavenumber ($\omega$ is angular frequency and $c$ is soundspeed).

Let us assume the sound field to be generated by the source field
$\mu(\theta,\phi,k)$ on a sphere of radius $r'$:
\begin{align}
\hspace{-2em}p(r,\theta,\phi,k) =  \int_{d\Omega} \mu(\theta',\phi',k) G(x,x',k) 
\sin(\theta') r'^2 d\theta' d\phi',
\end{align}
where $G(x,x',k)$ is a Green's
function and $x=(r,\theta,\phi)$. 

The Green's function $G(x,x',k)$ can be written as
\begin{align}
G(x,&x',k) =  
\frac{\mathrm{e}^{-j k \|  x - x'\|}}{4\pi \|x-x'\|}  \nonumber\\ 
 = &\sum_{n=0}^\infty \sum_{m=-n}^n  
(-j)\, k\, h^{(2)}_n(kr') j_n(kr) Y_n^{-m}(\theta',\phi')
  Y_n^{m}(\theta,\phi) \,
\nonumber\\ 
  &  \,\mathrm{for} \,r'  \geq r
\end{align}
where $h_n^{(2)}$ is the spherical Hankel function of the second
kind.

Let us define the source field at radius $r'$  by a discrete sequence of spherical harmonics coefficients:
\begin{align}
\mu(\theta',\phi',k) &= 
\sum_{n=0}^\infty \sum_{m=-n}^n  \gamma_n^m(k)  Y_n^m(\theta',\phi').  \label{q:mugamma}
\end{align}

Integrating $\mu(\theta',\phi',k) G(x,x',k) $ over the 2-sphere of radius $r'$, using
orthogonality of the spherical harmonics, we obtain an expression for
$p(r,\theta,\phi,k)$ in terms of $\gamma_n^m(k)$ that facilitates mode matching.  This
relates the sound field \eqref{q:basic} with the source field on the 2-sphere
\eqref{q:mugamma}: 
\begin{align}
\gamma_n^m(k) 
&= B_n^m(k) r'
  \,j^{-n}\,\mathrm{e}^{jkr'}, \quad kr' \rightarrow \infty
\label{q:gammabeta}
\end{align}
where we used the asymptotic behavior of
$h_n^{(2)}$ \cite{spors2012,williams1999fourier}: $
\lim_{kr \rightarrow \infty} h^{(2)}_n (k r) = j^{(n+1)} \frac{\mathrm{e}^{-jkr}}{kr}$.

\eqref{q:gammabeta} is the main result of this section. It shows that emphasizing
the source field on the sphere does \textit{not} correspond to a straight emphasizing of the
sound field $p(r,\theta,\phi,k)$. 

The source field \eqref{q:mugamma} in the frequency domain is
\begin{align}
\mu(\theta,\phi,k) 
= &r' \mathrm{e}^{jkr'} \sum_{n=0}^\infty \sum_{m=-n}^n g_n B_n^m(k)  Y_n^{m}(\theta,\phi) , \nonumber\\
&\quad k r' \rightarrow \infty ,
\label{q:driving3}
\end{align}
where we defined, for later convenience, $g_n = (-j)^{n} $. Except for a radius-dependent
scaling, the vector $g_n$ provides the mapping from the ambisonics coefficients to the
spherical harmonics representation of the source field. 

As we also aim to derive time-domain emphasis operators, we note that by applying the
inverse Fourier transform $\frac{1}{2\pi} \int \,\cdot\,\,  \mathrm{e}^{j\omega
  t}d\omega$ \eqref{q:driving3} can also be written in the time domain. 


\subsection{Emphasizing the Angular Dependency of a Signal}
\label{s:emphasis}
Our objective in this section is to enable us to emphasize a
particular direction with low computational complexity. That is, we aim to
emphasize (``sharpen'') the angular dependency of the source field $\mu$ associated with
the sound field.  To reduce computational requirements we aim to find expressions for the
ambisonics coefficients $B_m^n(\cdot)$ of the sharpened sound field without explicitly
calculating the source field $\mu$.

Our emphasis operator uses two time scales. The first time scale
resolves the temporal behavior of the monaural source signals and is characterized by
their bandwidth. The second time scale captures the rate of change 
of the parameters of the emphasis operator. For a time-invariant
emphasis in a particular direction, these parameters do not change in
time. More commonly, the second time scale resolves  
the changes in the spatial arrangement and loudness of the sound sources.
Frequency domain implementations in practice use
time-frequency transforms. Hence both frequency and time domain implementations can
accomodate time-dependencies on the second time scale.

 We first define the emphasis operator for the source field
$\mu(\theta,\phi,k)$. We start with a suitable function $v(\theta,\phi,k): [0,\pi] \times
[0,2\pi] \rightarrow [0,\infty)$ that is real and, ideally, nonnegative and can be used to
emphasize the source field $\mu(\cdot,\cdot,k)$ over the 2-sphere. 
The emphasis operation is then 
\begin{align}
\tilde{\mu}(\theta,\phi,k)  =  v(\theta,\phi,k) \,  \mu(\theta,\phi,k).
\label{q:moddriv}
\end{align}
For the  pressure $p$, the emphasis operation is not a
multiplication. We define $\nu$ as a general emphasis operator that also applies to
pressure and is the multiplication with $v$ \eqref{q:moddriv} in the source-field domain.  



We simplify our notation by introducing a single index
for the spherical harmonics and omitting function arguments where that is not ambiguous. Let
$\tilde{Q}$ be the degree 
of the ambisonics expansion for $\mu$. We
define $Q=(\tilde{Q}+1)^2$. Assuming that the source field
$\mu$ is of finite degree we have
\begin{align}
\mu &=  \sum_{q=0}^{Q-1} \gamma_q  Y_q.
\label{q:pmu}
\end{align}
We choose $v$ to be of degree $\tilde{L}$ and define $L=(\tilde{L}+1)^2$:
\begin{align}
v &=  \sum_{l=0}^{L-1}  V_l  Y_l .
\label{q:pv1}
\end{align}
Note that the finite degree $\tilde{L}$ prevents strict nonnegativity.

Exploiting that the spherical harmonics form a basis of the 2-sphere, we can write each
multiplication of pairs of spherical  harmonics as a weighted sum of spherical
harmonics. Let $Y^{(Q)}(\theta,\phi)$ be the $Q$-dimensional column vector 
$[Y_0(\theta,\phi),  Y_1(\theta,\phi), \cdots , Y_{Q-1}(\theta,\phi), ]^T$.
Let us denote the Kronecker product with $\otimes$.  We furthermore use that the
multiplications of two spherical harmonics of degree $L$ and $Q$ can be written as a
weighted sum of spherical harmonics with degree less or equal to $Q+L$. Thus, we can write 
\begin{align}
Y^{(Q)} \otimes  Y^{(L)}  = C \,  Y^{(P)},
\label{q:Clebsch}
\end{align}
where $C \in\mathrm{R}^{QL\times P}$ with $\tilde{P}=\tilde{Q}+\tilde{L}$ and
$P=(\tilde{P}+1)^2$ is a real, non-square matrix with
Clebsch-Gordan coefficients as elements. The matrix $C$ depends only on the degree of
the ambisonics representation and on the degree of the emphasis operator $v$. Hence it can
generally be computed off-line.

The standard formula for the multiplication of spherical harmonics shows that the
matrix $C$ is sparse and this can be exploited. However, as will be shown below, for
static or slowly varying emphasis (static or slowly varying emphasis operator), optimal
computational efficiency can be obtained without consideration of the sparsity of $C$. 

We can use standard formulas for the Clebsch-Gordan coefficients to compute
$C$. However, given that relation \eqref{q:Clebsch} exists, we can use it to
compute the matrix of Clebsch-Gordan coefficients ($C$) by creating a set of linear
equations corresponding to a set of random (or selected) angles. 

The emphasized source field $\tilde{\mu}$ can be written in terms of the
spherical harmonics expansions for $\mu$ and $v$:
\begin{align}
\tilde{\mu} = \nu \mu = v \mu &=  \sum_{q=0}^{Q-1}  \sum_{l=0}^{L-1}  \gamma_q V_l  Y_q Y_l \\
& =  \sum_{i=0}^{P-1} Y_i \, (C^T \,( \gamma^{(Q)} \otimes V^{(L)} ))_i,
\label{q:pv2}
\end{align}
where we used \eqref{q:Clebsch}. We now have the ambisonics expansion of $\tilde{\mu}$ in
terms of ambisonics expansions for $\mu$ and $v$. 

Let $\circ$ be the Hadamard (element-wise) product and $g^{(P)} = [g_{n(0)}, \cdots g_{n(P-1)}]^T$ , where
 with some abuse of notation, $n(i) =\lfloor \sqrt{i} \rfloor$ is the degree $n$ in $Y_n^k = Y_i$. 
It then follows from \eqref{q:gammabeta} that \eqref{q:pv2} implies that the emphasis operator for
the ambisonics coefficients of a pressure field satisfies
\begin{align}
g^{(P)} \circ \tilde{B}^{(P)} =  C^T \left( (g^{(Q)} \circ  B^{(Q)}) \otimes V^{(L)} \right),
\label{q:pv4}
\end{align}
which specifies the degree-expanded ambisonics
representation of the sound field \eqref{q:basic} after emphasis.

Next, we discuss the efficient computation of \eqref{q:pv4}. We will show that if the
emphasis operator $V$ is
time-invariant, then \eqref{q:pv4} can be computed with one $P\times Q$ matrix multiply
per sample, requiring $PQ$ multiplies to compute all output channels. Thus, for a degree-1
ambisonics representation, $\tilde{Q}=1$, and a degree-2 emphasis operator, $\tilde{L}=2$,  only four
multiplies per output channel are required.

One approach to obtaining high computational efficiency for computing \eqref{q:pv4} is to
exploit that $C$ and $V$ are fixed or slowly varying. Let $1^{(Q)}=[1,\cdots,1]^T$ be a
$Q$-dimensional vector of ones. Some algebra leads to
\begin{align}
g^{(P)} \circ \tilde{B}^{(P)}
             & =  \bar{C}^T  \,   ((g^{(Q)} \circ B^{(Q)}) \otimes 1^{(L)}  ),
\label{q:pv6}
\end{align}
where we wrote $\bar{C}^T =  C^T \circ (1^{(P)}( V^{(L)} \otimes 1^{(Q)})^T) $, which is a
matrix that retains the dimensionality of $C^T$.  Finally we note that we can define a matrix
 $A^{(QL)}= I^{(Q)} \otimes 1^{(L)}$, where $I^{(Q)}$ is the identity matrix with $Q$ rows and
 columns, such that $ B^{(Q)} \otimes 1^{(L)} = A^{(LQ)} B^{(Q)}$.  Thus, we can write
\begin{align}
\tilde{B}^{(P)} & =  \left( \mathrm{diag}^{-1}(g^{(P)})\,\bar{C}^T  \,  A^{(LQ)} \mathrm{diag}(g^{(Q)})\right)  B^{(Q)}  .
\label{q:superfast}
\end{align}

As  $\left( \mathrm{diag}^{-1}(g^{(P)})\,\bar{C}^T  \,  A^{(LQ)}
  \mathrm{diag}(g^{(Q)} ) \right)  \in \mathbb{C}^{P\times Q}$ is 
a time-invariant matrix for a fixed emphasis operator and slowly time-varying for an adaptive
emphasis operator, it be computed off-line or at a slow update rate. Thus, we
have proven that $PQ$ multiplies for each sample suffice for the emphasis
operator.

The usage of \eqref{q:pv4} without further modification is also relevant,  as
it facilitates rapid adaptation of the emphasis operator $\nu$. In this second approach, the
emphasis operator  
is composed of two components: \textit{i}) a Kronecker product operation, which
is an unrolled outer product of a $Q\times 1$ signal vector and an $L\times 1$ emphasis vector,
followed by \textit{ii}{) a $P \times QL $ matrix multiply.  While the size of
the matrix $C^T$ is larger than that of $\bar{C}^T  \,  A^{(LQ)}  $ in
\eqref{q:superfast}, it is a sparse matrix. From the explicit formula for the
Clebsch-Gordan series for product of two spherical harmonics formulas it follows
that the number of multiplies is also for this case $PQ$.  While the formulation \eqref{q:pv4} is less
conveniently structured, the fact that it has no computational overhead may
make it more attractive for scenarios that require rapid updates.

For both approaches discussed, the emphasis operation can be performed in the 
time domain or in the time-frequency domain. The domains result in different outcomes.
The  methods apply to real and complex spherical harmonics expansions.

\subsection{An Adaptive Emphasis Operator}
\label{s:finding}

The emphasis operator can be used to place an acoustic spotlight on a particular direction,
in the ambisonics domain. For a time-invariant and source-independent emphasis the tools
defined in section \ref{s:emphasis} suffice. However, a natural application of the 
emphasis operator is to emphasize an existing source-field power
distribution over directions. This section discusses how to find such an \textit{adaptive}
emphasis operator $\nu$. In most applications, the required adaptation rate
is low making both emphasis approaches of section \ref{s:emphasis} relevant.
 
Considering the pressure $p$ as a stochastic
process, a design for $v(\theta,\phi,k)$ with the desired emphasis result is: 
\begin{align}
 v(\theta,\phi,k)  & = \beta \, \mathrm{E}[  | \mu(\theta,\phi,k)   |^\alpha ],
\label{q:vdesign1}
\end{align}
where $\mathrm{E}$ is ensemble expectation, $\beta$ is a normalization and $\alpha > 0$ is a real
constant. A time-domain representation can also be used. As the time-domain representation
averages over frequencies, the results are not the same. 

Even integer values for $\alpha$ result in tractable expressions for \eqref{q:vdesign1}.
We illustrate the case $\alpha=2$.  Emphasis strengths can be varied by repeating the procedure and
by using a lower degree ambisonics representation as basis.
To evaluate \eqref{q:vdesign1} for $\alpha=2$, we first rewrite the complex conjugate of
the source field as  
\begin{align}
\mu^*(\theta,\phi,k) 
 = & r' \mathrm{e}^{-jkr'} \sum_{n=0}^\infty \sum_{m=-n}^n j^{n}
\breve{B}_n^{m}(k)   Y_n^{m}(\theta,\phi) , \nonumber\\
&\quad k r' \rightarrow \infty .
\label{q:drivingconj3}
\end{align}
where we used $Y_n^{m*} = Y_{n}^{-m}$ and defined $\breve{B}_n^{m}(k) = B_n^{-m*}(k)$.


Next,  we again simplify the notation and write \eqref{q:driving3} using a single index
and without the function arguments: 
\begin{align}
\mu(\theta,\phi,k) = & r' \mathrm{e}^{jkr'}\sum_{q=0}^\infty  g_{n(q)} B_q \,  Y_q,
&\quad k r' \rightarrow \infty ,
\label{q:driving5}
\end{align}
where, with some abuse of notation, we write $n$ as a function of $q$.
Based on \eqref{q:drivingconj3} and \eqref{q:driving5} we can rewrite \eqref{q:vdesign1}  as
\begin{align}
 v(\theta,\phi,k)
& = \beta \, r^{'2} \sum_{l=0}^\infty  \sum_{q=0}^\infty g_{n(l)} g_{n(q)}^* \mathrm{E}[ B_l 
\breve{B}_q ]  \, Y_l  Y_q .
\label{q:power2}
\end{align}
The same form is also obtained for the time-domain case.


We rewrite \eqref{q:power2} as an expansion in
spherical harmonics rather than products of spherical harmonics. We assume that the
original source field
$\mu$ is a degree-$\tilde{Q}$ source field and use $Q=(\tilde{Q}+1)^2$. Selecting the
normalization $\beta =\frac{\beta'}{r^2}$:
\begin{align}
\frac{v(\theta,\phi, k)}{\beta'} &= \frac{1}{r^{'2}} \mathrm{E}[ | \mu(\theta,\phi,k) |^2
                      ] \nonumber\\
& = \sum_{i=0}^{P-1} Y_i \, (C^T \, 
\mathrm{E}[ (g^{(Q)} \circ B^{(Q)}) \otimes (g^{(Q)*} \circ \breve{B}^{(Q)}])])_i  \nonumber\\
\label{q:power3}
\end{align}
where $P=(2\tilde{Q}+1)^2$, and $C$ is of the form of \eqref{q:Clebsch}.   \eqref{q:power3}
provides the adaptive emphasis operator in the desired form of an expansion in 
the spherical harmonics $Y_i$.

In a practical application the expectation must be approximated.  It is natural to assume
ergodicity for the signals and approximate the expectation operator with an averaging over
time in each bin of a time-frequency representation and simply over time in a time-domain
representation. 

The expectation contributes most to the computational effort for
\eqref{q:power3}. However, the averaging can be undersampled to satisfy any computational
complexity constraint. The remaining 
computations in \eqref{q:power3} are done at the update rate of the emphasis operator,
which typically is low. Hence these remaining computations normally do not play a
significant role in the computational complexity of finding $v$. The sparsity of $C$ can
be exploited to minimize computational effort.

In general, the emphasis operator changes the sound field also in the sweet zone
where the sound field computed from the unemphasized representation is
accurate. In the adaptive case, emphasis in this region is usually undesirable. The problem can
be removed by using a projection onto the nearest solution for which the sweet zone
is unchanged \cite{Kleijn2017g}.  Because of the orthogonality of the spherical harmonics,
the projection can be implemented by overwriting the low-degree ambisonics coefficients
with the corresponding original coefficients and requires no additional computational
effort. 

\section{Results}

The aim of this letter is to show that static and adaptive directional emphasis can be implemented
at negligible computational complexity in the ambisonics domain. For perceptual experiments that
show the benefit of directional emphasis we refer to other work: \cite{Wabnitz11,
  Wabnitz12} and in particular \cite{Kleijn2017g}, which implements adaptive emphasis
in the source-field domain. In this section, we illustrate the operation of the static and 
adaptive emphasis operator. All computations were performed in the spherical
harmonic domain with the methods of section \ref{s:theory} using complex spherical
harmonics. For illustration only, the
results were converted to the shown densities on the 2-sphere.  

 Fig. \ref{f:splotlighta01} shows mean source fields and the enhancement
operator $v$  (the acoustic emphasis operator) on the 2-sphere for simulated sound fields. On the
left is a degree-2 signal enhanced by a static degree-4 ambisonics acoustic
emphasis operator.  Only the signal highlighted by the emphasis operator is clearly audible. On the  
right we show the behavior of an adaptive acoustic emphasis operator, for
a degree-2 signal enhanced by a degree-8 ambisonics adaptive emphasis operator ($\alpha=4$).
As expected, negative values for $v$ in the source domain were small and away
from the high-intensity areas. Their significance reduces further with increasing emphasis operator degree, and
increasing emphasis operator smoothness.

\begin{figure}[t]
  \centering
  \includegraphics[width=0.235\textwidth]{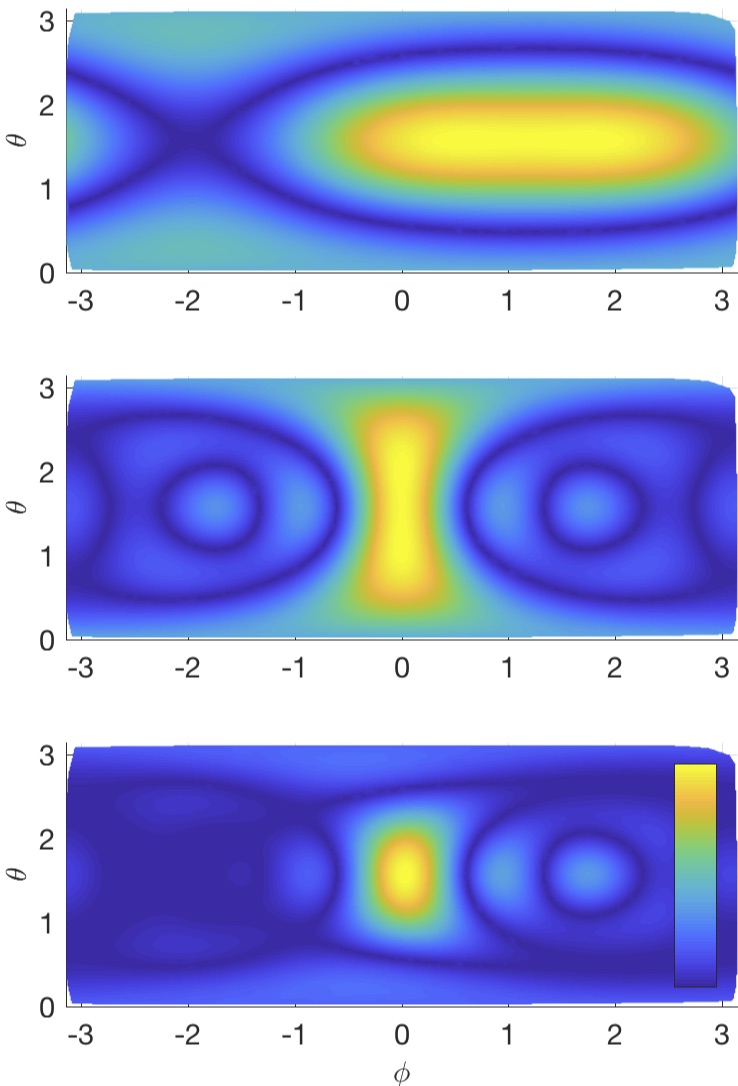}
  \includegraphics[width=0.235\textwidth]{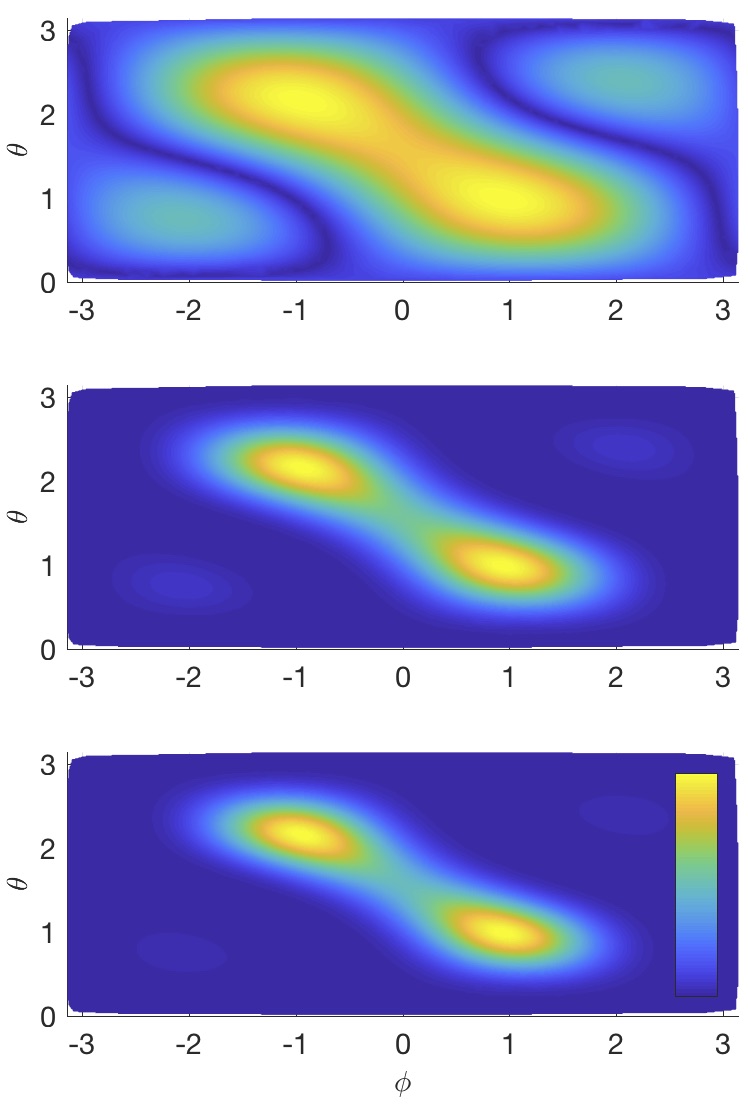}
  \caption{Mean absolute source field on the 2-sphere for degree-2 signals (top), a degree-4
    static emphasis operator (middle left) and an adaptive degree-8 emphasis operator (middle right), and the resulting
    emphasized signals (bottom). The color bars in the bottom figures show color
    linearly proportional to distance along the vertical axis, increasing upward.
  \label{f:splotlighta01}}
\end{figure}

\section{Conclusion}
\label{s:conclusion}
\vspace{-0.3em}
Practical implementations of ambisonics truncate its series representation of the
soundfield because of constraints on estimation and bit rate. For standard rendering, the
consequence of the truncation is that the timbre and directionality
of the acoustic scenario, as perceived by the listener, are distorted. A strengthening of
the directionality of the ambisonics representation can address these problems
\cite{Wabnitz11,  Wabnitz12, Kleijn2017g}.

We have shown that it is possible to define an emphasis operator that strengthens
the directionality of the sound field at negligible computational cost by using
Clebsch-Gordan coefficients. In contrast
to existing idempotent methods \cite{Wabnitz11,  Wabnitz12, Kleijn2017g}, the procedure
is attractive for real-time implementation and is particularly suitable for rendering over
headsets. More-over it facilitates a static emphasis. 

The new method can be applied to time domain or time-frequency
domain ambisonics representations. It can be used for representations based on real and
complex spherical harmonics (only the latter was illustrated). 

\bibliographystyle{IEEEtran}
\bibliography{AmbisonicsRefs,BeamFormerRefs}

\end{document}